\documentstyle[preprint,aps,epsfig]{revtex}
\begin{document}

\draft

\newcommand{\be}{\begin{equation}}
\newcommand{\ee}{\end{equation}}
\newcommand{\bea}{\begin{eqnarray}}
\newcommand{\eea}{\end{eqnarray}}
\newcommand{\lp}{\frac{1}{16 \pi^2}}
\newcommand{\ms}{m_{\rm soft}}
\newcommand{\mw}{m_{\rm W}}
\newcommand{\mi}{m_{\rm I}}
\newcommand{\mg}{M_{\rm GUT}}
\newcommand{\ler}{\stackrel{\scriptstyle <}{\scriptstyle\sim}}
\newcommand{\ger}{\stackrel{\scriptstyle >}{\scriptstyle\sim}}
\newcommand{\lag}{\langle}
\newcommand{\rag}{\rangle}
\newcommand{\nn}{\nonumber}
%%%%%%%%%%%%%%%%%%%%%%%%%%%%%%%%%%%%%%%%%%%%%%%%%%%%%%%%%%%%%%%%
%     Abbreviated journal names
%%%%%%%%%%%%%%%%%%%%%%%%%%%%%%%%%%%%%%%%%%%%%%%%%%%%%%%%%%%%%%%%
\def\npb#1#2#3{    {\it Nucl. Phys. }{\bf B #1} (19#2) #3}
\def\plb#1#2#3{    {\it Phys. Lett. }{\bf B #1} (19#2) #3}
\def\prd#1#2#3{    {\it Phys. Rev. }{\bf D #1} (19#2) #3}
\def\prep#1#2#3{   {\it Phys. Rep. }{\bf #1} (19#2) #3}
\def\prl#1#2#3{    {\it Phys. Rev. Lett. }{\bf #1} (19#2) #3}

\def\de{{\rm det}(\lambda_u \lambda_d)}
\def\dh{{\rm det}(\lambda_u \lambda_d \lambda_h)}
\def\ap{\approx}
\def\a{\alpha}
\def\l{\lambda}
\def\n{N_g}
\def\r{\rho}
\def\c{\cdot}
\def\t{\rm{1 TeV}}
\def\be{\begin{equation}}
\def\ee{\end{equation}}
\def\bea{\begin{eqnarray}}
\def\eea{\end{eqnarray}}

\tightenlines

\preprint{\begin{tabular}{r} KAIST-TH 98/18
\\  hep-ph/9809286 \end{tabular}}

\title{Small Instanton Contribution to the Axion Potential 
in Supersymmetric Models}

\author{
Kiwoon Choi\thanks{E-mail: kchoi@higgs.kaist.ac.kr} and 
Hyungdo Kim\thanks{E-mail: hdkim@supy.kaist.ac.kr}
}

\address{Department of Physics, Korea Advanced Institute of Science
and Technology \\
Taejon 305-701, Korea \\
}

\maketitle

\begin{abstract}

Small size QCD instantons may spoil the axion solution
to the strong CP problem if  QCD is not asymptotically free at 
high energy scales.
We examine this issue in supersymmetric models using a manifestly
supersymmetric scheme to compute the 
axion potential induced by small size instantons. 
Applying this scheme for a class of illustrative models, 
it is found that the resulting high energy axion potential is
highly model--dependent, but 
suppressed  by more powers of the soft supersymmetry breaking
parameters and/or of the other small mass scales  
than what is expected based on a naive instanton graph analysis.
Our analysis suggests that the axion solution is stable against
the small QCD instanton effects in a wide class of
supersymmetric models even when QCD is not asymtotically free at high energy
scales.

\end{abstract}

\pacs{PACS Number(s): }
%\newpage

%\bigskip

\section*{\bf 1. Introduction}

One of the most attractive solutions to the strong CP problem \cite{strongCP}
is to introduce a spontaneously broken $U(1)_{PQ}$
symmetry whose {\it explicit} breaking is entirely
given by the QCD anomaly \cite{PQ}.
Once such a global $U(1)_{PQ}$ is assumed,
it is usually taken for granted that $\theta_{QCD}$ is dynamically relaxed
down to a sufficiently small value by the vacuum expectation value (VEV)
of the associated 
pseudo-Goldstone boson, the axion.
However there are some class of models that this may not be 
true. If QCD is {\it not} asymptotically free at energy scales above the weak 
scale, small size QCD instantons may induce a significant
high energy axion potential $V_{\rm HE}$ whose minimum does {\it not} coincide
with the minimum of the conventional  low energy axion potential $V_{\rm LE}$.
If $V_{\rm HE}\ger 10^{-9}V_{\rm LE}$,
the dynamical relaxation mechanism of $\theta_{QCD}$ would be spoiled 
and then one looses
the original motivation to  introduce $U(1)_{PQ}$.

Perhaps the most interesting class of 
models in which the effects of small size instantons can be sizable
are supersymmetric models.
Some  supersymmetric Grand Unified Theories (GUTs) 
like $E_6$ and also many supersymmetric models 
derived from superstring theories
contain extra colored multiplets
which would dramatically change
the QCD $\beta$ function at energy scales above the weak scale, even 
making it 
positive in some cases. 
In fact,  small size instanton contribution to the axion potential
in models with extra colored particles has been studied before  
\cite{dine,flynn}
based on a simple dimensional analysis for instanton graphs.
However, when applied for supersymmetric models, this scheme
can overlook a possible cancellation between different instanton
graphs which is due to supersymmetry (SUSY),
thereby yielding a highly overestimated axion potential.
To avoid this difficulty, 
in this paper we introduce a manifestly supersymmetric scheme 
to compute the axion potential
and apply it for a class of illustrative models.
Although an explicit analysis is made only for a limited
class of models and also  
the resulting  axion potential is highly  
model-dependent, our analysis  suggests that
the effects of small size instantons in supersymmetric models
are  much more suppressed 
than what is expected based on 
a naive instanton graph analysis.
This  implies
that the axion solution to the
strong CP problem is stable against the effects of small instantons
in a wide class of SUSY
models even when QCD is not asymtotically free at high
energy scales.

To proceed, let us discuss in more detail how the dynamical
relaxation of $\theta_{QCD}$ can be spoiled by  
the high energy axion potential induced by small size QCD instantons.
At the axion scale $f_a$ where
$U(1)_{PQ}$ is spontaneously broken, the axion effective lagrangian
is given   by
\be
{\cal L}_{\rm axion}=\frac{1}{2}(\partial_{\mu}a)^2+
\frac{1}{f_a}\partial_{\mu}a J^{\mu}+\frac{1}{32\pi^2}
\frac{a}{f_a}F\tilde{F},
\ee
where  $J^{\mu}$ denotes a generic current in the model, and $F$ and 
$\tilde{F}$ are the gluon field strength and its dual, respectively.
Integrating out the degrees of freedom at energy
scales above the weak scale $M_W$,
the nonderivative axion coupling to the QCD anomaly, i.e.
$aF\tilde{F}$, would induce
a high energy axion potential $V_{\rm HE}$ through the
QCD instantons with size
$\r\ll M_W^{-1}$:
\be
V_{\rm HE}=-V_0\cos(a/f_a-\gamma).
\ee
Both the size of this axion potential
($V_0$), and the location of the minimum
($\gamma$),
depend upon the details
of the physics at high  energy scales  around $\r^{-1}$.
Unlike $V_{\rm HE}$,
the low energy axion potential induced by the dynamics below the weak scale
can be unambiguously computed.  Applying  the chiral
perturbation theory for the low energy QCD instanton amplitudes,
one finds \cite{strongCP}
\be
V_{\rm LE}=-f_{\pi}^2m_{\pi}^2
\sqrt{m_u^2+m_d^2+2m_um_d\cos(a/f_a)}/(m_u+m_d),
\ee
where $m_u$ and $m_d$ denote the current masses
of the up and down quarks, $f_{\pi}=94$ MeV and $m_{\pi}$ are the pion decay
constant and mass, respectively.
Here the axion VEV
($\langle a/f_a\rangle$) 
is defined to
correspond to the QCD vacuum angle
which is required to be less than $10^{-9}$ from the non-observation
of the neutron electric dipole moment:
\be
\theta_{QCD}=\langle a/f_a\rangle \ler 10^{-9}.
\ee
The axion VEV is then determined by the total axion potential
$V_{\rm axion}=V_{\rm LE}+V_{\rm HE}$.
Obviously, in order for the axion VEV to satisfy the above
phenomenological bound,
the high energy axion potential has to be suppressed as 
$V_0/f_{\pi}^2m_{\pi}^2 \ler 10^{-9} \gamma^{-1}$.
If one wishes to achieve CP-conserving
QCD  {\it independently of} the details of CP violation
at  high energy scales,
i.e. independently of $\gamma$, one would need
\bea
\delta\equiv \frac{V_{\rm HE}}{ 10^{-9}f_{\pi}^2m_{\pi}^2} \ler 1. 
\label{strong}
\eea
In fact, there is a  negligibly small extra low energy
potential \cite{georgi}
$$\delta V_{\rm LE}\simeq 10^{-14}f_{\pi}^2m_{\pi}^2
\sin \delta_{KM}\sin (a/f_a)
$$ 
due to the low energy dynamics
involving the CP violation by the Kobayashi-Maskawa phase $\delta_{KM}$.
This would shift the axion VEV also, but the resulting shift 
$\langle\delta a/f_a\rangle\simeq 10^{-14}\sin\delta_{KM}$
is well within the experimental bound.

In the next section,  we describe how the instanton-induced
$V_{\rm HE}$ can be computed  in generic 
supersymmetric models in a manifestly supersymmetric way.
In sect. 3, we apply this method
for a class of illustrative SUSY models
including extra quark multiplets
and make a numerical estimate of the largest possible $V_{\rm HE}$.
Based on this analysis, we   conclude in sect. 4  
that the axion solution to the
strong CP problem is stable against the effects of small instantons
in a wide class of SUSY
models even when QCD is not asymtotically free at high
energy scales.

\section*{\bf 2. 
Small instanton induced axion potential in SUSY models}

In this section, we discuss how the high energy axion potential induced
by small size QCD instantons can be computed in a manifestly
supersymmetric way.
Generic supersymmetric models contain a SUSY breaking sector which
would provide a dynamical seed for SUSY breaking.
The resulting  SUSY breaking is  transmitted to the
observable sector through some messenger interactions characterized
by the messenger scale $M_m$.
Integrating out the fields in SUSY breaking sector and also
in the messenger sector, one is left with an effective lagrangian
of the observable sector fields in which 
SUSY appears to be softly broken.
This effective lagrangian  describes the observable
sector dynamics at energy scales below  $M_m$.
The size of the messenger scale
$M_m$ depends upon the character of messenger interactions.
In gravity-mediated models \cite{grav}, $M_m$ is given by the Planck scale
$M_P$, while in gauge-mediated models \cite{gauge}
$M_m$ is somewhat model-dependent
but  typically far below $M_P$. 
At any rate, soft SUSY breaking
in the observable sector is operative only at energy scales
below $M_m$. As a result, QCD instantons with size $\rho\ler M_m^{-1}$ 
do not contribute to the axion potential due to SUSY fermion zero
modes, e.g. gluino zero modes.
To be definite, here we concentrate on gravity-mediated models,
however our study
can be easily applied for gauge-mediated models also.

Our starting point is 
the effective lagrangian of observable sector fields at  energy
scales around $\rho^{-1}$ where
$\rho$ denotes the size of the small QCD instantons:
\be
{\cal L}  = \int d^2 {\theta}d^2 {\bar{\theta}} \, 
{\eta}{\Phi_i}^{\dagger} {\Phi_i} + 
\frac{1}{32 {\pi}^2} [ \int d^2 {\theta} \, Y W^{\alpha} W_{\alpha} + h.c. ] 
+ [ \int d^2 {\theta} W ( \Phi_i, Z ) + h.c. ].
\label{lag}
\ee
Here $\Phi_i$ denote generic chiral superfields in the model,
$W_{\alpha}$ are the spinorial chiral superfields
for the gauge multiplets, 
$\eta$ is a  real spurion superfield whose
auxiliary $D$-component corresponds to the soft scalar mass, and
finally  $Y$ and $Z$ stand for the chiral spurions whose
auxiliary $F$-components correspond to the gaugino mass
and the soft $A$ (or $B$) coefficients, respectively.
Explicitly, we have
\bea
\eta &=& 1+m_0^2\theta^2\bar{\theta}^2, \nonumber \\
Y &=& \frac{8\pi^2}{g^2}+i\theta_{QCD}+16\pi^2 m_{1/2}\theta^2,
\nonumber \\
Z &=& \{Z_{\lambda}=(1+A\theta^2)\lambda ,\quad 
Z_{\mu}=(1+B\theta^2)\mu\},
\label{spurion}
\eea
where $\lambda$ and $\mu$ denote the trilinear Yukawa couplings
and the bilinear $\mu$-parameter in the superpotential.
It is assumed that all soft
breaking terms are described by the universal soft scalar mass
$m_0$, the universal gaugino mass $m_{1/2}$, and the universal trilinear
$A$ and bilinear $B$ coefficients,  which are presumed
to have the weak scale values.
The size of the small instanton-induced axion potential 
is {\it not} sensitive to the detailed
form of the soft parameters and 
thus its order of magnitude can be estimated  
in the scheme assuming a universal form of
soft parameters.

Once the effective lagrangian 
at energy scales  around $\r^{-1}$ is given as (\ref{lag}),  
the low energy interactions 
induced by small QCD instantons   
can be summarized by \cite{choi} 
\bea
{\cal L}_{\rm ins}
 &&= \int d^2 {\theta}d^2 {\bar{\theta}} \, 
[e^{-n_1Y}K_{\rm eff} ( \Phi_i, \Phi_i^*; D_{\alpha},
\bar{D}_{\alpha}; \bar{Z, Z^*; \eta}) +h.c.] \nonumber \\
&&+ [ \int d^2 {\theta} e^{-n_2Y}W_{\rm eff} ( \Phi_i, Z) + h.c. ],
\label{leff}
\eea
where $n_1$ and $n_2$ are positive integers corresponding
to the instanton winding number and $D_{\alpha}$ denotes the supercovariant
derivative acting on either superfields or spurions.
In order  to generate an axion
potential,  all  external fields with vanishing VEVs, e.g.
external quark superfields, 
in the instanton amplitudes have to be integrated out 
to induce a term including 
{\it only} the superfields with  nonvanishing VEVs.
Then the  axion potential can be derived from the terms
in $W_{\rm eff}$ or $K_{\rm eff}$ 
by replacing all the superfields by their VEVs,
the spurion $Y$ by $Y+ia/f_a$, and finally integrating out the
Grassmann coordinates $\theta$ and $\bar{\theta}$.
Note that the supercovariant derivatives are not allowed 
to appear in $W_{\rm eff}$ since $\bar{D}\Phi=\bar{D}Z=0$
and also  $D^2 \Phi$ ($D^2 Z$) is not a chiral superfield
(spurion).

Let $m_{\rm soft}$ collectively denote the soft mass parameters, i.e.
$$
\{ m_{\rm soft} \}=\{m_0, m_{1/2}, A, B \}.
$$
Usually any nonzero VEV of the auxiliary $F$-component of
the observable sector field $\Phi_i$ is due to the presence
of soft SUSY breaking terms and thus it
is suppressed by some powers (typically a single power)
of $m_{\rm soft}$, e.g.
\be
\langle \Phi_i \rangle\ap (1+m_{\rm soft}\theta^2) \langle
\phi_i \rangle \ ,
\label{superfield}
\ee
where $\phi_i$ is the scalar component of the superfield $\Phi_i$.
One simple consequence
of this observation is that the axion potential
arising from
$K_{\rm eff}$ is suppressed  at least by two powers of
$\r m_{\rm soft}$ since it is obtained after the Grassmann integration
over $d^2\theta d^2\bar{\theta}$, while the axion potential 
from  $W_{\rm eff}$ is suppressed 
at least by a single power of $\r m_{\rm soft}$.

The instanton-induced effective K\"ahler potential  $K_{\rm eff}$ and 
the effective superpotential $W_{\rm eff}$  can be systematically
computed by expanding them in powers of 
the small dimensionless quantities:
$\r\Phi_i$, $\rho D_{\alpha}D^{\alpha}$,
$Z_{\lambda}$, and $\rho Z_{\mu}$.
In fact, such an expansion is severely constrained by the selection rules
dictated  by the internal symmetries of the tree level 
superpotential $W(\Phi_i,Z)$ in (\ref{lag}) under which both 
the superfields and spurions transform.
Another useful requirement is that
it should have a sensible limiting behavior
when some of the couplings are turned off.
As we will see, together with the holomorphy,
these requirements {\it forbid} 
$W_{\rm eff}$ to include a term which would 
contribute  to the axion potential by replacing the superfields
by their VEVs.
As a result, 
{\it the leading contribution to the axion
potential comes from the effective K\"ahler potential $K_{\rm eff}$ 
and thus is suppressed at least by two powers of
$\r \ms$}.

Once the superfields in $K_{\rm eff}$ are replaced  by their VEVs and 
also the Grassmann integration is performed, the expansion
in powers of $\r \Phi_i$, $\r D^{\alpha}D_{\alpha}$, $\r Z_{\mu}$,
and $Z_{\lambda}$ appear to be an expansion in   
$\langle \r \phi_i\rangle$,
$\r m_{\rm soft}$, $\r \mu$ 
and also the Yukawa couplings
in the model.
Generically the field VEVs are calculable in terms of
$m_{\rm soft}$ and/or the other mass parameters, e.g. $\mu$ and $M_P$,
in the model. 
One can then easily identify 
the dominant term
by counting the powers of $\r m_{\rm soft}$ and $\r\mu$
(and also of the Yukawa couplings if necessary)
in the resulting axion potential.
Although the selection rules 
determine how the  
high energy axion potential depends upon 
the various parameters and the VEVs,
they do not fix the additional suppression
factor arising from the loops in the instanton
graph.
The axion potential induced by small
instantons is essentially given by closing the fermionic zero modes
in the instanton-induced  multiple fermion operator
in the dilute gas approximation.
Once the dominant  term 
in the expansion of $K_{\rm eff}$ were identified,
the corresponding instanton graph can be  
identified also. This allows us to 
estimate the loop suppression factor by
counting the number of loops 
in the graph, thereby
completing the estimate
of the size of the high energy axion potential.

The above procedure computing the instanton-induced axion
potential  is based on a manifestly supersymmetric
formulation and thus automatically takes into account 
the possible cancellation dictated by SUSY.
Anticipating SUSY cancellation
between different instanton graphs would be highly nontrivial
in a naive diagrammatic analysis.
In fact, in many cases
one can draw an instanton graph which is consistent with the selection rules
and the sensible limiting behavior,
but with a less powers of
$\r m_{\rm soft}$
than those required by SUSY. This implies that
such graph is cancelled
by other instanton graphs  which are related to the original graph
by SUSY. 
Our method automatically avoids such  misleading graphs since it includes
only the instanton graphs allowed by both  SUSY and the 
selection rules.
Following the recipe described above,
in the next section we will compute 
the small instanton-induced axion potential
in a class of illustrative SUSY models.

\section*{\bf 3. 
Computation and numerical analysis}

There are many possible supersymmetric extensions of the Standard Model,
starting from the minimal supersymmetric standard model (MSSM)
to other models including additional matter multiplets.
Here we are mainly interested in models with additional
colored multiplets which would change the QCD $\beta$ function
at high energy scales,
thereby leading to a sizable
high energy axion potential
generated by  small QCD instantons.
To examine this possibility, in this
section  we consider a model  containing $N_h$-flavors
of isosinglet quark and antiquark multiplets
($h+h^c$)
%which form a vector-like representation 
%of the standard model gauge group 
and also a gauge-singlet
superfield ($S$)  {\it in addition to} the 3 generations of
the MSSM quark multiplets (isodoublet quarks $Q$ and
isosinglet antiquarks $U^c+D^c$).
This model is a rather simple generalization of the MSSM, but
still shows many of  the essential features
of the axion potential induced by small instantons in generic
SUSY models.

The tree level superpotential of our model includes 
\be
W  = Z_u Q H_u U^c + Z_d Q H_d D^c + Z_h \frac{S^{k+1}}{{M_P}^k} h h^c
     + Z_1 \frac{S^{n+1}}{{M_P}^n} H_u H_d + Z_2 \frac{S^{l+3}}{{M_P}^l}
\ , 
\label{superpotential}
\ee
where  $l$, $k$, $n$ are {\it non-negative} integers,
and $H_u$ and $H_d$ denote the isodoublet Higgs superfields.
Terms involving lepton superfields are irrelevant for the analysis of 
the axion potential and thus are ignored.
Here $Z_{u,d}$ are 3$\times$3 matrices, $Z_h$ is $N_h\times N_h$
matrix, and all couplings are spurions which are determined by
the Yukawa couplings and the universal $A$-coefficient:
\bea
&& Z_u = {\lambda}_u ( 1 + A {\theta}^2 ) \ , \quad
Z_d = {\lambda}_d ( 1 + A {\theta}^2 ) \nn \\
&& Z_1 = {\lambda}_1 ( 1 + A {\theta}^2 ) \ , \quad
Z_2 = {\lambda}_2 ( 1 + A {\theta}^2 ) \ , \\
&&Z_h = {\lambda}_h ( 1 + A {\theta}^2 ).
\eea
A characteristic feature of the models considered here
is that 
all mass scales  are determined
in terms of the two basic mass scales $\ms$ and $M_P$ and also
the three non-negative integers $(l,k,n)$.
(See Eqs. (14) and (15).) 
It is straightforward to repeat our following  analysis 
for other type of
models including a mass scale which is not related with
$\ms$ and $M_P$.

To proceed, let us first consider the ground state configuration
of the model. We assume that $S$ has a sizable renormalizable
Yukawa coupling,
for instance the Yukawa coupling $\lambda_h$ 
with $h+h^c$ in the case of $k=0$, or a renormalizable Yukawa coupling
with lepton-like superfields in other cases with
$k > 0$.
This would make the running
soft mass of $S$ to become  {\it negative} at energy scales
around $\langle S\rangle$.
Then the effective potential of $S$ is schematically given by 
\be
V  \ap -\ms^2 S^2 + \frac{S^{2(l+2)}}{{M_P}^{2l}},
\ee
where we have omitted the coefficients of order unity.
The resulting VEV of $S$ is
\be
\langle S\rangle \ap M_P  \left(\frac{\ms}{M_P}\right)^{\frac{1}{l+1}},
\label{svev}
\ee
yielding  the supersymmetric masses  ($\mu_H$ and
$\mu_h$) of $H_u+H_d$ 
and $h+h^c$ as follows:
\bea
&& \mu_H = \lambda_1 {\langle S\rangle^{n+1}\over {M_P}^n} \ap
\ms \left(\ms\over {M_P}\right)^{n-l\over l+1}, \nonumber \\
&& \mu_h = \lambda_h {\langle S\rangle^{k+1}\over {M_P}^k} \ap
\ms  \left(\ms \over {M_P}\right)^{k-l\over l+1},
\label{mh}
\eea
where we have assumed that $\lambda_1, \lambda_2$ and $\lambda_h$ 
are of order unity.
These  masses can {\it not} be significantly
smaller than 
$\ms$ which is presumed to be of order  the weak scale,
and thus
the non-negative integers $l,k,n$ are required to satisfy
\be
n \leq l \, , \quad k\leq l.
\label{con1}
\ee

For the case with $n=l$,
the Higgsino mass $\mu_H$ is of order $\ms$, and then
$H_u$ and $H_d$ can be interpreted as the MSSM Higgs doublets 
which have the weak scale VEVs generating the masses of $W$, $Z$, 
and also of  the quarks and leptons:
\be
\langle H_u\rangle\ap \langle H_d\rangle\ap \ms \quad
{\rm for} \quad n=l.
\label{h1}
\ee
In this case, 
the observed quark mass spectrum gives 
\be
{\rm det}(\lambda_u\lambda_d)\ap 10^{-16} \quad
{\rm for} \quad n=l.
\ee
However for $l>n$, we have $\mu_H\gg \ms$
and then $H_u$ and $H_d$ should be interpreted as additional
massive Higgs doublets {\it other than} the MSSM Higgs doublets.
Such additional (superheavy) Higgs doublets 
appear quite often in SUSY $E_6$  models and also string orbifold models
\cite{munoz}.
At any rate, for $\mu_H\gg \ms$, $H_u$ and $H_d$ have vanishing
VEVs:
\be
\langle H_u\rangle=\langle H_d\rangle=0 \quad
{\rm for} \quad l>n,
\label{h2}
\ee
and then $\lambda_u$ and $\lambda_d$
are {\it not} constrained at all.
As a final remark on the vacuum configuration,
one can explicitly confirm that
\be
\langle F_i\rangle=
\left\langle \frac{\partial W}{\partial \phi_i}
\right\rangle
\ap \ms \langle \phi_i\rangle,
\ee
for the models under consideration,
which would assure the expression (\ref{superfield}) for the superfield VEVs.

Given the above described  vacuum configuration,
the $N_h$ flavors of
additional isosinglet quarks ($h+h^c$) affect
the QCD $\beta$ function at 
energy scales above $\mu_h$, particularly make it {\it positive}
if $N_h >3$.
Obviously we are then interested in small instantons with
$\r \ler \mu_h^{-1}$.
To be definite,
we will consider here only   the three
distinctive values of $N_h$:
the case with one $h+h^c$ 
in each generation ($N_h=3$), the case
with  two $h+h^c$ in each generation
($N_h=6$),
and the case with three $h+h^c$ in each generation ($N_h=9$).
We then require that the QCD fine structure constant
$\alpha$ does not blow up at energy scales below
$M_{GUT} = 2 \times 10^{16} {\rm GeV}$ for $N_h=6$ and $N_h=9$
which is a conservative condition.
We do not require no blowing up of $\alpha$
above $M_{GUT}$ to $M_P$ because there is a possibility
that enlarged gauge group at $M_{GUT}$ can change the running
behavior of $\alpha$. 
This leads to the following conditions on $\mu_h$ and thus
on $k$ and $l$:
\bea
& N_h=6: &\mu_h \gtrsim  10^{-5}(\ms M_{GUT})^{1/2}
\rightarrow   \frac{k+1}{l+1} \lesssim \frac{9}{10},
\nonumber \\
& N_h=9: &\mu_h \gtrsim  5\times 10^{-4}(\ms M_{GUT}^2)^{1/3}
\rightarrow       \frac{k+1}{l+1} \lesssim \frac{29}{45}.
\label{con2}
\eea
Within the range $l \le 2$ which we will concentrate on, 
all possible sets of $(l,k)$ are restricted
to {$(1,0)$,$(2,0)$,$(2,1)$} for $N_h=6$ and {$(1,0)$,$(2,0)$} for $N_h=9$
by the above constraints.

As we will see, the axion potential induced by small instantons
is highly sensitive to the Yukawa 
couplings  $\lambda_i$ ($i=u,d,1,2,h$).
For $\lambda_1$,
$\lambda_2$, and $\lambda_h$ which are unknown in any case, 
we simply assume that they
are all of order unity.
Note that this corresponds  to a conservative
choice when applied to  check whether the resulting
high energy axion potential satisfies the strong
CP condition (\ref{strong}).
About $\lambda_u$ and $\lambda_d$,
we have two possibilities.
For the case with $n=l$, $H_u$ and $H_d$ are the
usual MSSM Higgs doublets and thus
${\rm det}(\lambda_u\lambda_d)\ap 10^{-16}$.
However for $l>n$, 
$H_u$ and $H_d$ are additional (superheavy)
Higgs doublets with vanishing VEVs,
and then
$\lambda_u$ and $\lambda_d$ can be assumed to be of order unity
once again as a conservative choice.

Small instanton-induced axion potential has the tunnelling factor
$e^{-\frac{2\pi}{\alpha(\rho)}}$ 
for the single instanton contribution ($n_1=n_2=1$) which gives
the most dominant effect.
This tunnelling factor can be determined by the 
one-loop $\beta$ function:
\be
\beta(E)=E\frac{dg}{dE} = -b_0(E) \frac{g^3}{16 {\pi}^2},
\ee
where $b_0(E)=3$ for $E\lesssim \mu_h$, but $b_0(E)=3-N_h$ for
$E\gtrsim \mu_h$.
We then have
\bea
\exp (-2 \pi /\alpha(\r)) &\ap&
(\ms/\mu_h)^3
\left(\r \mu_h\right)^{3-N_h}\exp (-2\pi/\alpha (m_{\rm soft}))
\nonumber \\
&\ap& 2 \times 10^{-31} (\r m_{\rm soft})^{3-N_h}  
(\ms/M_P)^{\frac{N_h (l-k)}{l+1}},
\label{instanton}
\eea
where we have used the expression (\ref{mh}) for $\mu_h$
together with the numerical value of 
$\exp (-2 \pi/\alpha(\ms)) \ap 2\times 10^{-31}$
for $\ms\ap 1$ TeV.

Another important ingredients in computing
the instanton-induced axion potential
are the selection rules dictated by the underlying
theory  at high  energy scales around $\r^{-1}$.
>From the tree level superpotential, 
we can find various global symmetries under which
both the superfields and the spurions transform. 
Low energy effective theory obtained by integrating out 
high momentum modes should preserve these symmetries.
The global symmetries of the tree level superpotential
(\ref{superpotential}) include
\bea
G &=& SU(3)_Q \times SU(3)_{U^c} \times SU(3)_{D^c} 
\times SU(N_h)_h \times SU(N_h)_{h^c}
\nonumber \\
&& \times U(1)_A
\times U(1)_X \times U(1)_{X^{\prime}} \times U(1)_R \,
\eea
where 
$SU(N)_{\Phi_i}$  denotes the $SU(N)$ rotation
of the $N$-flavors of $\Phi_i$, and
the $U(1)$-charges of all superfields and spurions 
are depicted in Table 1.

Let us consider a
$G$-invariant effective superpotential
$e^{-Y}W_{\rm eff}$ with a 
sensible limiting behavior, which
would lead to an axion potential once the superfields
are replaced by their VEVs.
For a small instanton size $\r\lesssim \langle S\rangle^{-1}$,
it takes the following form:
\be
e^{-Y}W_{\rm eff} \sim \r^{-3}e^{-Y} {\rm{det}}( Z_u Z_d {\tilde Z_h}) 
{\tilde Z_1}^{N_1}
(\r^2 H_u H_d)^{N_2} (\r S)^{N_3} {\tilde Z_2}^{N_4} \ ,
\ee
where 
\be
{\tilde Z_h}=\frac{Z_h}{{(\r M_P)}^k},
\quad
{\tilde Z_1}=\frac{Z_1}{{(\r M_P)}^n},
\quad
{\tilde Z_2}=\frac{Z_2}{{(\r M_P)}^l}.
\label{dimensionless}
\ee
The integers $N_i$ ($i=1,2,3,4$) in the above are required to be
{\it non-negative} to have a sensible limiting behavior.
This condition can {\it not} be compatible with the $U(1)_R$-selection
rule 
\be
N_1 + N_4 +2 = 0,
\ee
and thus {\it no such an effective superpotential
is allowed.}

Since it does not have to be holomorphic,
a generic instanton-induced K\"ahler potential
which would lead to an axion potential
can be written as
\be
e^{-Y}K_{\rm eff} = \r^{-2} e^{-Y}
{\rm{det}} (Z_u Z_d {\tilde Z_h}) {{\tilde Z_1} \choose {\tilde Z_1}^*}^{N_1}
{\r^2 H_u H_d \choose \r^2 H_u^* H_d^*}^{N_2} {\r S \choose \r S^*}^{N_3} 
{{\tilde Z_2} \choose {\tilde Z_2}^*}^{N_4}
{\r D_{\alpha}D^{\alpha} \choose \r \bar{D}_{\alpha}\bar{D}^{\alpha}}^{N_5} 
F(\eta) \ ,
\label{kahler0}
\ee
where $F(\eta)$ is a function of the real spurion $\eta=1+m_0^2\theta^2
\bar{\theta}^2$ and the supercovariant derivatives
are understood to be applied for either the superfields or
the spurions.
The $G$-selection rules 
require first of all the factor
det($Z_uZ_d{\tilde Z_h}$).
One may then insert either $N_1$-powers of $Z_1$ or its complex
conjugate, $N_2$-powers of $H_uH_d$ or its
complex conjugate, and so on, to make $K_{\rm eff}$ 
to be fully $G$-invariant.
To make the notation simpler,
in the following we will use the notation
in which
a negative power corresponds
to the positive power of the complex conjugated  fields or spurions.
In this notation, the above
K{\"a}hler potential can be written as
\be
e^{-Y} K_{\rm eff}= \r^{-2} e^{-Y} {\rm{det}} (Z_u Z_d {\tilde Z_h}) 
{\tilde Z_1}^{N_1}
(\r^2 H_u H_d)^{N_2} (\r S)^{N_3} {\tilde Z_2}^{N_4} (\r D^2)^{N_5} F(\eta),
\label{kahler}
\ee
where now $\{N_i\}$ can be  negative integers
but still are constrained by the selection rules of 
$U(1)_X$, $U(1)_{X^{\prime}}$ and $U(1)_R$ as follows:
\bea
N_2&=&N_1+3  ,\nn \\
N_3&=&(n+1)N_1+(l+3)N_4+(k+1)N_h ,
\label{ncon3}\\
N_5&=&N_1+N_4+3. \nn
\eea

In fact, there can be two additional suppression factors 
that are encoded in $F(\eta)$.
The first one is the loop suppression factor which is always there,
while the second one appears only when $\r\lesssim f_a^{-1}$.
Let $\phi_{PQ}$ denotes the scalar field responsible
for the spontaneous $U(1)_{PQ}$-breaking. We then have
\be
\langle \phi_{PQ}\rangle=f_a e^{ia/f_aN_{DW}},
\ee
where the positive integer $N_{DW}$ corresponds to
the axion domain wall number.
Then the $U(1)_{PQ}$-selection rule,
i.e. the invariance under 
\bea
\phi_{PQ} &\rightarrow& e^{i \alpha} \phi_{PQ} ,\nonumber \\
\theta_{QCD} &\rightarrow& \theta_{QCD} - \alpha N_{DW},
\eea
 requires that 
the axion potential induced by a very small instanton 
with $\r\lesssim f_a^{-1}$ includes 
the factor $(\r\langle\phi_{PQ}\rangle)^{N_{DW}} $.
This factor is replaced by the simple phase factor
$e^{ia/f_a}$ for larger instantons with $\r\gtrsim f_a^{-1}$.
Summarizing these, we have
\be
F(\eta=1) \ap 
\epsilon_a\left(\frac{1}{4\pi^2}\right)^L 
\label{f}
\ee
where 
\bea
\epsilon_a &\ap& \left\{
\begin{array}{ll}
(\r f_a)^{N_{DW}} \quad &{\rm for} \quad
\r\lesssim f_a^{-1}, \\
1 \quad &{\rm for} \quad \r\gtrsim f_a^{-1}
\end{array}\right.
\eea
and $L$ denotes the number of loops in the instanton
graph.

After replacing the superfields in $K_{\rm eff}$
by their VEVs and also
integrating over the Grassmann variable,
we find
\bea
V_{\rm HE}(\r)
&\ap& 
\r^{-4} F(1) e^{-\frac{2\pi}{\a(\r)}} (\r \ms)^{2+|N_5|}
(\r M_P)^{-N_0}
(\r^2 \langle H_uH_d\rangle)^{|N_2|}
\nonumber \\
&\times& (\r \langle S\rangle)^{|N_3|} 
{\rm det}( \l_u \l_d \l_h) \l_1^{|N_1|} \l_2^{|N_4|},
\label{master1}
\eea
where 
\be
N_0=k N_h +n |N_1| +l |N_4|.
\ee
Note that the above axion potential includes
the {\it model-independent} SUSY suppression factor 
$(\r \ms)^2$ in addition to the tunnelling factor
$e^{-\frac{2\pi}{\alpha(\r)}}$ and also 
the model-dependent suppression 
factors due to small field VEVs and couplings.
Using the field VEVs (\ref{svev}), (\ref{h1}) and (\ref{h2}) together with
(\ref{instanton}) and (\ref{f}), we finally arrive at 
\bea
V_{\rm HE}(\r) &\ap & 
2\times 10^{-31} \epsilon_a 
\r^{-4}
(\r \ms)^{5+N_S-N_h} (\r M_P)^{-N_0+\frac{l}{l+1}|N_3|} \nonumber \\
&\times& 
\left(\frac{1}{4\pi^2}\right)^L
(\frac{\ms}{M_P})^{\frac{N_h (l-k)}{l+1}}
{\rm det}( \l_u \l_d \l_h) 
\l_1^{|N_1|} \l_2^{|N_4|} 
\label{master2}
\eea
where
\bea
N_S
&=& \left\{
\begin{array}{ll}
2|N_2|+\frac{1}{l+1}|N_3|+|N_5| \quad &{\rm for} \, \, l=n,
 \\ \\
\frac{1}{l+1}|N_3|+|N_5| \quad &{\rm for} \, \, l>n.
\end{array}\right.
\eea
Here  the loop number $L$ can be determined
by looking at an explicit instanton graph yielding
the above form of axion potential.
Note that 
$\langle H_u\rangle\ap \langle H_d\rangle \ap
\ms$ only for $l=n$.
For other cases with $l>n$, we have
$\langle H_u\rangle=\langle H_d\rangle=0$ and thus
only $N_2=0$ can yields a nonvanishing axion potential.

Given the values of $N_h=3,6,9$ and  $(n,l,k)$ satisfying
the constraints (\ref{con1}) and (\ref{con2}),
we can now pick out the values of $\{ N_i\}$ ($i=1\sim 5$)
which would give the most dominant 
contribution to $V_{\rm HE}$ for a fixed value of $\r$.
%while obeying the selection rules (\ref{ncon3}).
It turns out that for all cases studied here
such values of $\{N_i\}$ lead to a {\it negative}
total power of $\r$ in $V_{\rm HE}$
and thus  smaller instantons give a larger contribution.
Since we wish to see
whether $V_{\rm HE}$ is small enough to satisfy the strong CP condition 
(\ref{strong}), we are interested in
the largest possible value of $V_{\rm HE}$,
i.e. an upper bound.
We thus take $\r\ap 1/M_P$, being  the smallest possible
value of $\r$, and also
did {\it not} take into account the $f_a$-dependent suppression factor 
$\epsilon_a\ap (\r f_a)^{N_{DW}}$ in our numerical study.
It is then clear from eq. (\ref{master2}) that, 
for the fixed values of $(N_h, n,l,k)$ and $\r M_P\ap 1$, 
the largest possible value of $V_{\rm HE}$
is obtained for the minimal values  of $N_S$
and also of the loop number $L$.
In each case, it is rather straightforward
to find  the minimal  value of $N_S$
and also the corresponding instanton graph with
a  minimal number of loops.
(See Figures 2, 4, 5, 6.)
The results are summarized
in Table 2 for all cases with $l \le 2$.
It turns out that   $N_S <2 $ in all cases with $l=n \le 2$ and
thus $N_2=0$ even when $l=n$.
We stress that the values of $V_{\rm HE}$ in Table 2 are obtained
for  the smallest possible value of $\r$ ($\ap M_P^{-1}$)
and  the conservative choice of the
loop factor ($\ap \frac{1}{4\pi^2}$),
and also  the possible additional suppression factor $\epsilon_a
\ap (\r f_a)^{N_{DW}}$ for $f_a\lesssim \r^{-1}$ was not
taken into account.
In this sense, they could be largely overestimated
and thus
should be understood as a kind of upperbound on $V_{\rm HE}$.

As we have pointed out, a naive instanton graph analysis can yield
a misleading result with  {\it less} power of $\r \ms$
than what is required by SUSY and the internal symmetry selection rules.
Such a graph should be cancelled by other instanton graphs
which are related to each other by SUSY.
Avoiding this complication was the main motivation
to introduce a manifestly supersymmetric scheme to compute
the small instanton-induced axion potential.
To make the motivation more clear,
let us consider some examples of misleading instanton
amplitudes.
For  the case of $(n,l,k)=(0,1,0)$ and $N_h=9$,
there can be  a diagram like Fig. 1 which is consistent
with the selection rules of all internal symmetries,
but includes only one insertion of $\r\ms$
(the $A$ parameter in this case).
The size of the resulting amplitude is estimated
to be (for $\r^{-1}\ap f_a\ap M_P$)
\bea
V^{\rm mis}_{\rm HE} &\ap& (\frac{1}{4 \pi^2})^{12}
e^{-\frac{2 \pi}{\alpha(M_P)}}~ M_P ~{\rm det}(\l_u\l_d\l_h) 
{\l_1^*}^3 \l_2^* {\langle S \rangle}^2 A
\nonumber \\
&\ap& 6\times 10^{15} \, 
{\rm det}(\l_u\l_d\l_h) {\l_1^*}^3\l_2^* \, ~({\rm GeV})^4,
\eea
which would be much larger than $10^{-9}f_{\pi}^2m_{\pi}^2$
if all the Yukawa couplings are taken to be of order unity.
(Note that $l>n$ in this case and thus $H_u$ and $H_d$ 
should be interpreted as the superheavy Higgs doublets
which have a vanishing VEV and thus
can have a large Yukawa coupling with the standard model
quarks.)
Based on this result, one may conclude that the axion solution
to the strong CP problem can be spoiled by small size QCD instantons
in this model.
However to be consistent with both the  SUSY and the internal 
symmetry selection rules, the instanton amplitude
should include at least  two powers of $\r \ms$ and thus
the above amplitude should be cancelled by other instanton
amplitudes.
Our manifestly supersymmetric scheme tells us  
that the most dominant contribution 
comes from Fig. 2 including three insertions of
$\r\ms$. The resulting correct $V_{\rm HE}$
is then estimated to be
\bea
V_{\rm HE}
&\ap& (\frac{1}{4 \pi^2})^{11}
e^{-\frac{2 \pi}{\alpha(M_P)}}~ \frac{\ms^2}{M_P} ~{\rm det}(\l_u\l_d\l_h) 
{\l_1^*}^3 \l_2^* {\langle S \rangle}^2 A
\nonumber \\
&\ap& 6\times 10^{-14} \,
{\rm det}(\l_u\l_d\l_h) {\l_1^*}^3\l_2^* \, ~({\rm GeV})^4,
\eea 
which is smaller than  $10^{-9}f_{\pi}^2m_{\pi}^2$
even when all Yukawa couplings are taken to be of order unity.
In fact, in some cases one can draw an instanton graph
without any insertion of $\r\ms$.
For instance, for the case with $N_h=3$ and $n=l=k=0$,
Fig. 3 provides an instanton amplitude which is consistent
with all the internal symmetry selection rules
but does not contain any $\r\ms$.
Once again, such amplitude should be cancelled and
in the manifestly supersymmetric scheme 
the dominant contribution is from Fig. 4
including  two insertions of $\r\ms$.

%\bea
%V_{\rm HE} (\r \sim 1/M_P)  
%\ap 2 \times 10^{-31}
%(\frac{1}{16 \pi^2})^9 M_P^4 \l^{* 3} det(\l_u \l_d \l_h).
%\eea
%Figure 2 shows the corresponding diagram. 
%In the previous studies \cite{dine,flynn},
%it was argued that $\r m_{\tilde{g}} \simeq \r \ms$ suppression arises
%from the gluino zero modes, but in all the models studied here
%$V_{\rm HE}$ have {\it at least} two powers of SUSY suppressions
%and in some cases they are even further suppressed.

%Another interesting case is for $(n,l,k)=(0,1,0)$ with $N_h=9$.
%$(N_1,N_3,N_4,N_5)=(-3,0,-1,-1)$ can give the largest possible
%$V_{\rm HE}$.
%The suppression effect in $V_{\rm HE}$ due to supersymmetry 
%is given as $(\r \ms)^{2+|N_5|}=(\r \ms)^3$,
%and figure 3 shows the correct instanton diagram.
%Figure 4 shows one of misleading diagrams which has only one $\r \ms$
%suppression, and this is the diagram that we would expect naively by
%considering gluino mass insertion.
%We can not discriminate fake huge contributions which is required
%by supersymmetry to be canceled with other diagrams. 
%This may yield overestimation of $V_{\rm HE}$.
%The result (strong CP criterion $\delta$) is very sensitive
%to the suppression powers and if we estimate it with the misleading diagram
%(fig.4) we find $\delta > 1$.
%It is very important to know the correct suppression powers.

Let us finally  comment a technical point about the number
of loops in the instanton graph.
When two graphs have the same value of $N_S$, 
the dominant contribution comes from
the one with a smaller number 
of loops.
For instance, for the case with $N_h=9$ and
$(n,l,k)=(1,1,0)$, both  $(N_1,N_3,N_4,N_5)=(-3,3,0,0)$ 
and $(-3,-1,-1,-1)$ give the same value of $N_S=3/2$.
Fig. 5 shows the diagram of $(-3,3,0,0)$ 
with the number of loops $L=11$, while
Fig. 6 is for $(-3,-1,-1,-1)$ with the number of loops $L=13$.
In this example, Fig. 5 
has two more scalar VEVs ($\langle S\rangle$) 
than Fig. 6, but has two smaller number of loops.
Usually a graph with  more scalar VEVs
has a fewer scalar loops
because the number of scalar loops can be reduced by
closing the scalar field line by inserting its VEV.
Thus when there are two or more graphs with the same $N_S$,
the graph with a more scalar VEVs
has a fewer loops and thus gives a dominant contribution.

To close all the quark zero modes in the instanton
graph, one needs the insertion of 
either the bare quark mass or the Yukawa coupling involving a generic
complex scalar field $\phi$.
In our case, there is no bare quark mass and thus
all the quark zero modes are closed by 
the Yukawa couplings. As a result, the  instanton graph
involves many  $\phi$-lines 
which should be closed again to yield a vacuum amplitude.
There are many ways to close the $\phi$-lines.
One may close one $\phi$-line without yielding any 
scalar loop by the insertion of
$\langle \phi\rangle$,
or close two $\phi$-lines 
by inserting
a {\it complex} mass of the form 
$M_{\phi}^2\phi^2+{\rm h.c.}$ which would yield one
scalar loop.
Inserting the couplings involving higher powers of
$\phi$ to close the $\phi$-lines results in more scalar loops.
Obvioulsy the insertion of $\langle \phi\rangle$ in the
instanton graph
leads to the suppression by $\r\langle\phi\rangle$.
Also in supersymmetrc models, generically $M_{\phi}^2\ap \ms M_{\psi}$
where $M_{\psi}$ is a supersymmetric mass of the
fermionic partner $\psi$ of $\phi$, and thus
the insertion of $M_{\phi}^2$ leads to a suppression
by $\r\ms$.
Because of these suppressions,
although economical in reducing the scalar loops, 
instanton graphs in which most of the scalar lines
are closed either by $\langle \phi\rangle$ or  
$M_{\phi}^2$  do not give a dominant contribution
to $V_{\rm HE}$.
This is the reason why in our cases
the dominant contribution comes from the instanton
graph involving relatively many scalar loops.

%\bea
%L \sim N_f > N_f/2,
%\eea
%where $N_f$ is the total number of quark flavors.
%The number of loops $L \simeq N_f/2$ is expected if we close the fermionic
%zero modes by the insertion of scalar masses.
%However, there is no bare mass in our models
%and all the masses are dynamically generated by scalar VEVs, and
%the mass scales are related to $\ms$.
%After all, scalar fields can not be so heavy and the instanton diagrams
%which close fermionic zero modes by the insertion of scalar masses
%give highly suppressed $V_{\rm HE}$. 
%$S$ has very small mass $m_S^2 \simeq \ms^2$ 
%and mass insertion in the diagram gives $(\r \ms)^2$ suppression
%compared to the case in which zero modes are closed 
%by proper scalar interactions.
%$H_u$ and $H_d$ have also small masses compared to $M_P$,
%\bea
%(\r m_{H_u,H_d})^2 \simeq (\r M_P)^2 (\frac{\ms}{M_P})^{\frac{2(n+1)}{l+1}}, 
%\eea
%and
%at least $(\frac{\ms}{M_P})^{\frac{2}{3}} \sim 10^{-10}$ suppression
%happens if $H_u$ or $H_d$ mass is inserted for $l \le 2$.
%The diagrams which give the largest contributions to $V_{\rm HE}$
%do not include scalar masses and have more scalar loops $L \sim N_f$
%, and $V_{\rm HE}$ is more suppressed due to higher loop factors.

\section*{\bf 4. Conclusion}

In this paper, we have  studied the 
high energy axion potential $V_{\rm HE}$ induced by small QCD instantons
in supersymmetric models in which QCD is not
asymtotically free at high energy scales.
In such models,  $V_{\rm HE}$ may be larger than $10^{-9}f_{\pi}^2
m_{\pi}^2$ and then the axion solution to the strong CP
problem can not be successfully implemented.
To avoid the difficulty arising from the possible
SUSY cancellation between different instanton graphs,
we introduced a manifestly supersymmetric scheme to compute
$V_{\rm HE}$ in generic SUSY models and applied it for a
class of illustrative models.
In our scheme,  small instanton effects can be summarized
by the effective superpotential ($W_{\rm eff}$)
and also by the effective K\"ahler potential ($K_{\rm eff}$).
By imposing the selection rules and a sensible
limiting behavior together with the holomorphy, we showed
that $W_{\rm eff}$  does not allow a term which would contribute to
the axion potential and thus 
$V_{\rm HE}$ always comes from $K_{\rm eff}$.
As a simple consequence, $V_{\rm HE}$  is
suppressed {\it at least} by  two powers of $\r \ms$
where $\r$ denotes the small instanton size and $\ms$ is the soft
SUSY breaking mass which is presumed to be of order the weak
scale.
In addition to this model-independent suppression,
$V_{\rm HE}$ can be  suppressed further by  (i) small field
VEVs $\langle \r\phi_i\rangle$, (ii) small Yukawa couplings,
and also (iii) loop factors.
These model-dependent suppression factors are
carefully analyzed for a class of illustrative models
to estimate the possible largest value
of $V_{\rm HE}$ and the results are summarized in Table 2.
Our analysis suggests that
the axion solution to the
strong CP problem is stable against the effects of small instantons
in a wide class of SUSY
models even when QCD is not asymtotically free at high
energy scales.

\acknowledgements
HK would like to thank C. Munoz for helpful discussions.
This work is supported in part
by KOSEF Grant 981-0201-004-2, 
MOE Basic Science Institute Program BSRI-98-2434,
KOSEF through CTP of Seoul National University,
and KRF under the Distinguished Scholar Exchange Program.

\begin{table}[hbt]
\begin{center}
\caption{Quantum numbers of superfields and spurions}
\begin{tabular}{|c||ccccc|}
 &  $U(1)_A$ & $U(1)_X$ & $U(1)_{X^{\prime}}$ & $U(1)_R$&
\\
\hline
$Q$           & 1    & 0    & 0        & 1   & \\ \hline
$u^c,d^c$     & 1    & $-1$ & 0        & 1   & \\ \hline
$h,h^c$       & 1    & $0$  & 0        & 1   & \\ \hline
$H_u$, $H_d$  & 0    & 1    & 0        & 0   & \\ \hline
$S$           & 0    & 0    & 1        & 0   & \\ \hline
$e^{-Y}$      &12+2$N_h$& $-6$ & 0     & 6   & \\ \hline
$Z_u,Z_d$     & $-2$ & 0    & 0        & 0   & \\ \hline
$Z_h$         & $-2$ & 0    & $-(k+1)$ & 0   & \\ \hline
$Z_1$         & 0    & $-2$ & $-(n+1)$ & 2   & \\ \hline
$Z_2$         & 0    & 0    & $-(l+3)$ & 2   & \\ \hline
$d^2\theta$   & 0    & 0    & 0        & $-2$& \\ 
\end{tabular}
\end{center}
\label{table:1}
\vspace{7mm}
\end{table}

\vspace{5mm}
\begin{table}[b]
\begin{center}
\caption{The largest possible size of $V_{\rm HE}$
and $\delta=V_{\rm HE}/10^{-9}f_\pi^2m_{\pi}^2$ in each model.
If $\delta\protect\lesssim1$, the axion solution to the strong CP problem 
is untouched by small instantons.}
\begin{tabular}{||c||c|c|c||c|c|c|c||c||}
%\multicolumn{2}{|c||}
$N_h$ & $n$ & $l$ & $k$ &($N_1,N_3,N_4,N_5$)& $N_S$ & $L$(loop) 
& $V_{\rm HE}~({\rm GeV})^4$ & $\delta$ \\
\hline
3 & 0 & 0 & 0 & $(-3,0,0,0)$ & 0 & 8 &
$1.6 \times 10^{-17}$ & $4.1 \times 10^{-5}$ \\
3 & 0 & 1 & 0 & $(-3,0,0,0)$ & 0 & 8 &
$1.8 \times 10^{-24}$ & $4.5 \times 10^{-12}$\\
3 & 0 & 1 & 1 & $(-3,3,0,0)$ & 3/2 & 8 &
$1.8 \times 10^{-24}$ & $4.5 \times 10^{-12}$\\
3 & 0 & 2 & 0 & $(-3,0,0,0)$ & 0   & 8 &
$4.1 \times 10^{-32}$ & $1.0 \times 10^{-19}$\\
3 & 0 & 2 & 1 & $(-3,3,0,0)$ & 1   & 8 &
$4.1 \times 10^{-32}$ & $1.0 \times 10^{-19}$\\
3 & 0 & 2 & 2 & $(-3,1,-1,-1)$ & 4/3  & 12 &
$2.7 \times 10^{-28}$ & $6.6 \times 10^{-16}$\\
3 & 1 & 1 & 0 & $(-3,-3,0,0)$ & 3/2 & 8 &
$2.0 \times 10^{-63}$ & $5.1 \times 10^{-51}$\\
3 & 1 & 1 & 1 & $(-3,0,0,0)$ & 0 & 11 &
$2.6 \times 10^{-22}$ & $6.6 \times 10^{-10}$\\
3 & 1 & 2 & 0 & $(-3,-3,0,0)$ & 1 & 8 &
$2.0 \times 10^{-47}$ & $5.1 \times 10^{-35}$\\
3 & 1 & 2 & 1 & $(-3,0,0,0)$ & 0 & 11 &
$1.3 \times 10^{-21}$ & $3.3 \times 10^{-9}$\\
3 & 1 & 2 & 2 & $(-3,3,0,0)$ & 1 & 11 &
$1.3 \times 10^{-21}$ & $3.3 \times 10^{-9}$\\
3 & 2 & 2 & 0 & $(-3,-1,1,1)$ & 4/3 & 12 &
$6.6 \times 10^{-75}$ & $1.7 \times 10^{-62}$\\
3 & 2 & 2 & 1 & $(-3,-3,0,0)$ & 1 & 11 &
$6.6 \times 10^{-53}$ & $1.7 \times 10^{-40}$\\
3 & 2 & 2 & 2 & $(-3,0,0,0)$ & 0 & 14 &
$4.3 \times 10^{-27}$ & $1.1 \times 10^{-14}$\\
\hline
6 & 0 & 1 & 0 & $(-3,3,0,0)$ & 3/2 & 8 &
$1.8 \times 10^{-24}$ & $4.5 \times 10^{-12}$\\
6 & 0 & 2 & 0 & $(-3,3,0,0)$ & 1   & 8 &
$4.1 \times 10^{-32}$ & $1.0 \times 10^{-19}$\\
6 & 0 & 2 & 1 & $(-3,4,-1,-1)$ & 7/3   & 12 &
$2.7 \times 10^{-28}$ & $6.6 \times 10^{-16}$\\
6 & 1 & 1 & 0 & $(-3,0,0,0)$ & 0 & 11 &
$2.6 \times 10^{-22}$ & $6.6 \times 10^{-10}$\\
6 & 1 & 2 & 0 & $(-3,0,0,0)$ & 0 & 11 &
$1.3 \times 10^{-21}$ & $3.3 \times 10^{-9}$ \\
6 & 1 & 2 & 1 & $(-3,1,-1,-1)$ & 4/3 & 15 &
$8.6 \times 10^{-18}$ & $2.2 \times 10^{-5}$ \\
6 & 2 & 2 & 0 & $(-3,-3,0,0)$ & 1 & 11 &
$6.6 \times 10^{-53}$ & $1.7 \times 10^{-40}$ \\
6 & 2 & 2 & 1 & $(-3,-3,0,0)$ & 1 & 14 &
$4.3 \times 10^{-27}$ & $1.1 \times 10^{-14}$ \\
\hline
9 & 0 & 1 & 0 & $(-3,2,-1,-1)$ & 2 & 11 &
$5.9 \times 10^{-14}$ & $1.5 \times 10^{-1}$\\
9 & 0 & 2 & 0 & $(-3,1,-1,-1)$ & 4/3   & 12 &
$2.7 \times 10^{-28}$ & $6.6 \times 10^{-16}$\\
9 & 1 & 1 & 0 & $(-3,3,0,0)$ & 3/2 & 11 &
$2.6 \times 10^{-22}$ & $6.6 \times 10^{-10}$\\
9 & 1 & 2 & 0 & $(-3,3,0,0)$ & 1 & 11 &
$1.3 \times 10^{-21}$ & $3.3 \times 10^{-9}$ \\
9 & 2 & 2 & 0 & $(-3,0,0,0)$ & 0 & 14 &
$4.3 \times 10^{-27}$ & $1.1 \times 10^{-14}$\\
\end{tabular}
\end{center}
\label{table:2}
\end{table}
\vspace{5mm}

\begin{figure}[bht]
\caption{Instanton  graph which is consistent with all
the internal symmetry selection rules but is required to be cancelled by SUSY
for the case of
$N_h = 9$ and $(n,l,k)=(0,1,0)$. Here the solid lines with and without
waves denote the gluino and quark modes, respectively,
while the dotted lines are the scalar fluctuations of Higgs, squarks,
and $S$. The dark blobs represent the insertions of complex couplings
or field VEVs which are explicitly written in the graph.} 
\vspace{5mm}
\centerline{\epsfig{file=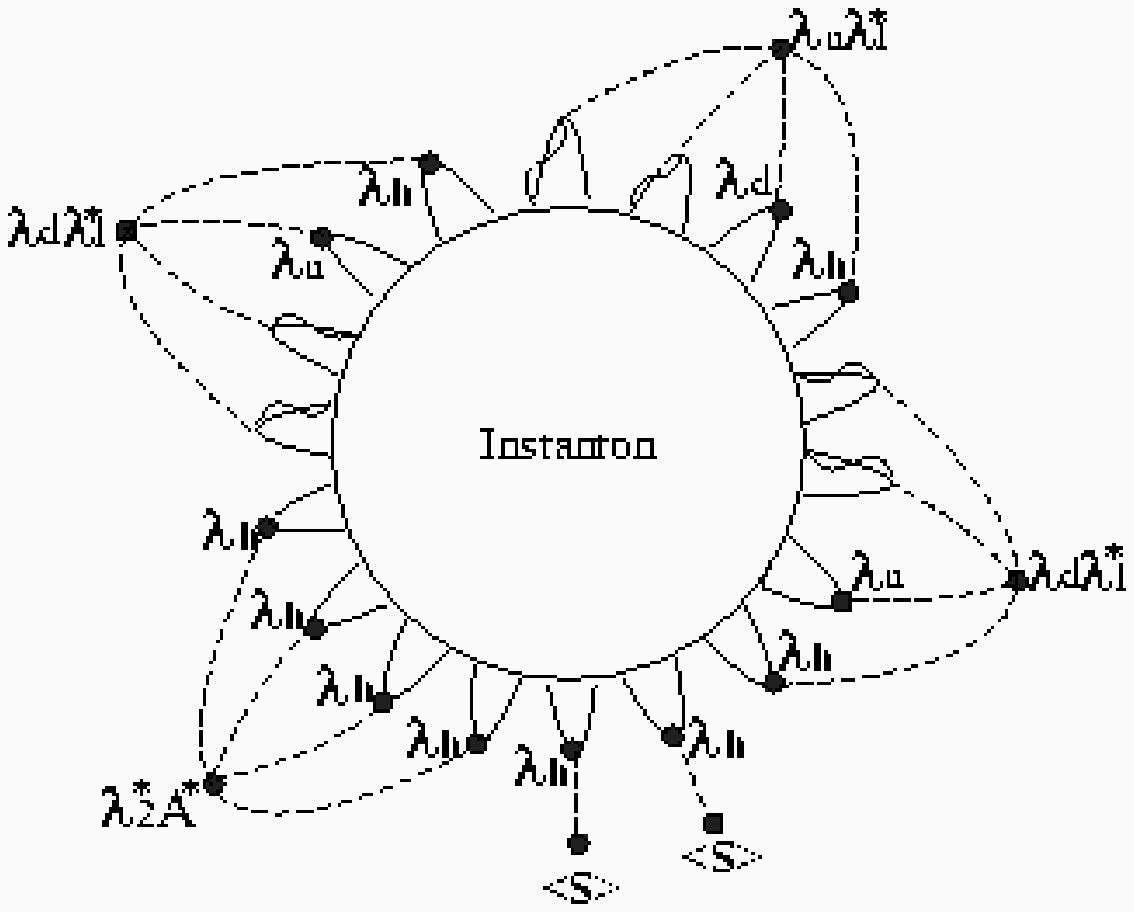, height=9cm}}
\label{fig4}
\caption{Dominant instanton graph 
for the case of $N_h = 9$ and $(n,l,k)=(0,1,0)$.} 
\vspace{5mm}
\centerline{\epsfig{file=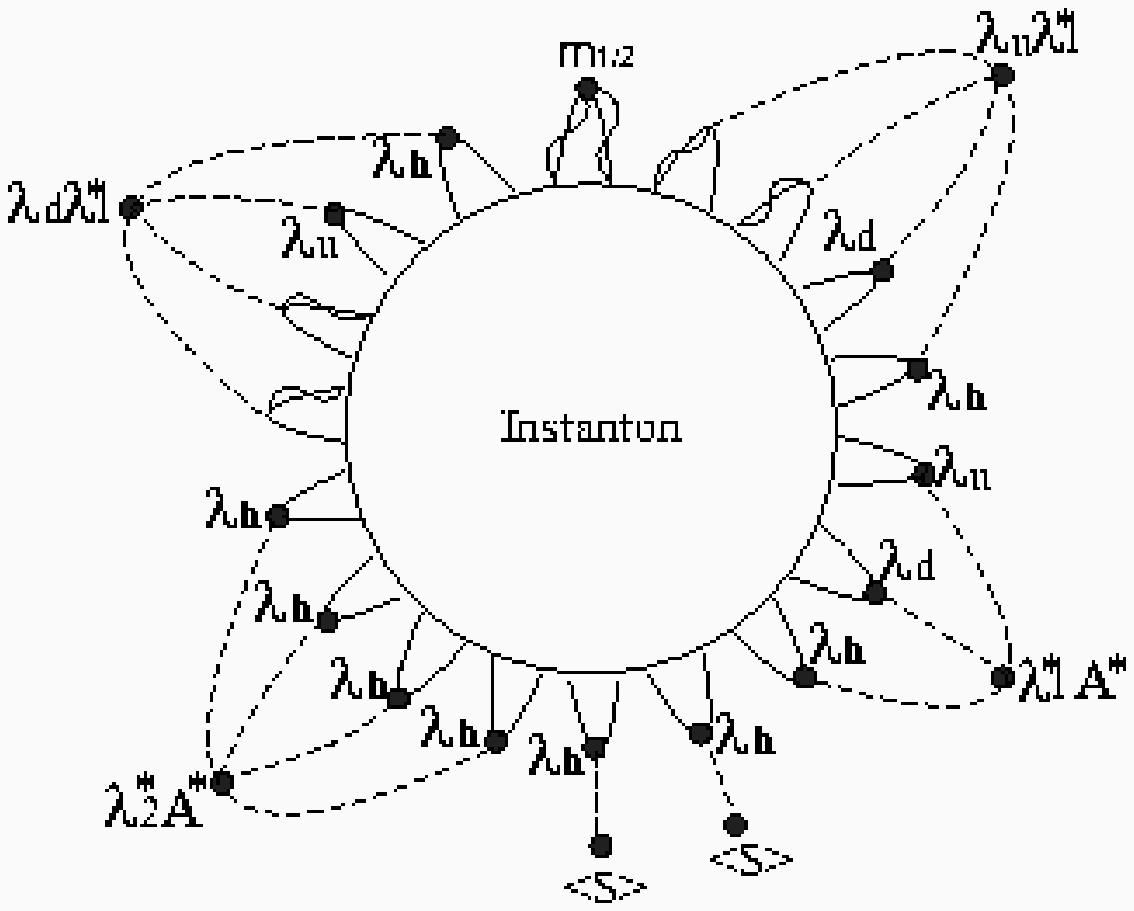, height=9cm}}
\label{fig3}
\newpage
\caption{Instanton graph which is  required to be cancelled by SUSY
for the case of $N_h = 3$ and $(n,l,k)=(0,0,0)$.} 
\vspace{5mm}
\centerline{\epsfig{file=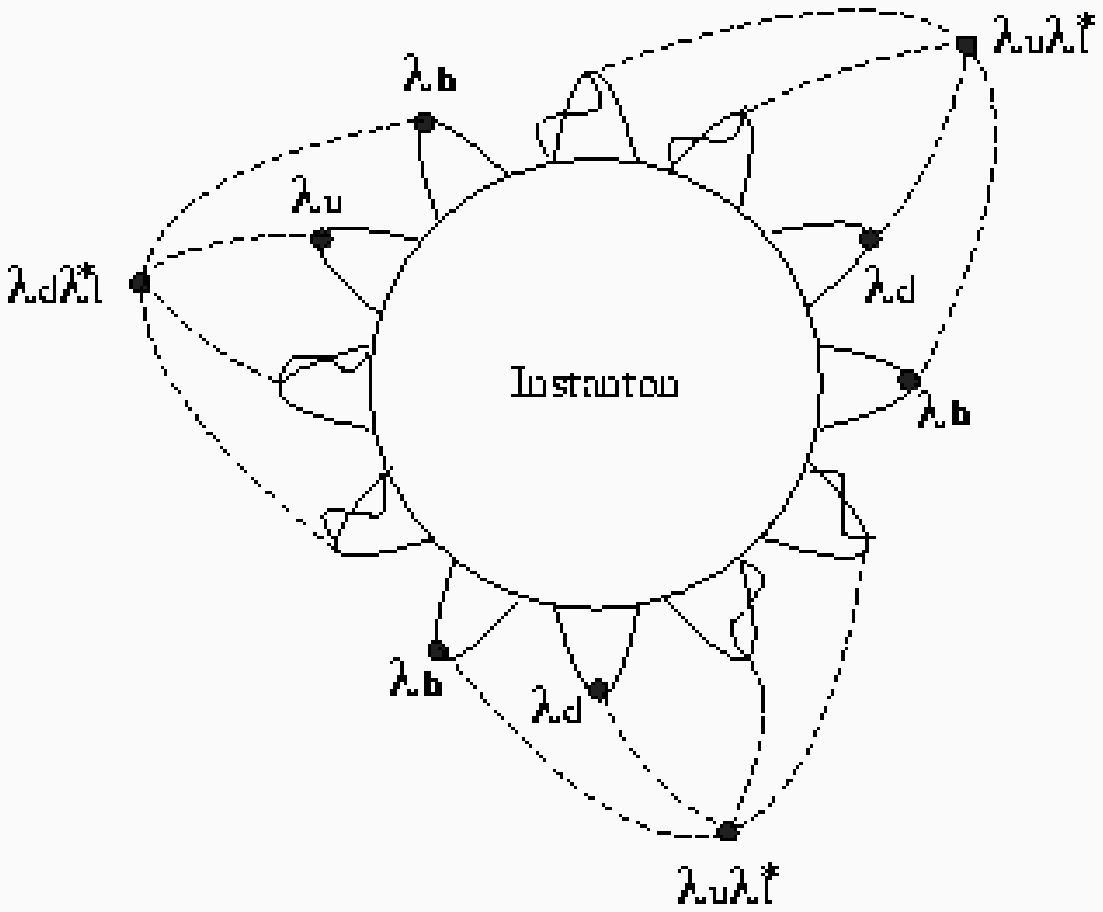, height=9cm}}
\label{fig2}
\vspace{10mm}
\caption{Dominant instaton graph for the case of $N_h = 3$ and 
$(n,l,k)=(0,0,0)$.} 
\vspace{5mm}
\centerline{\epsfig{file=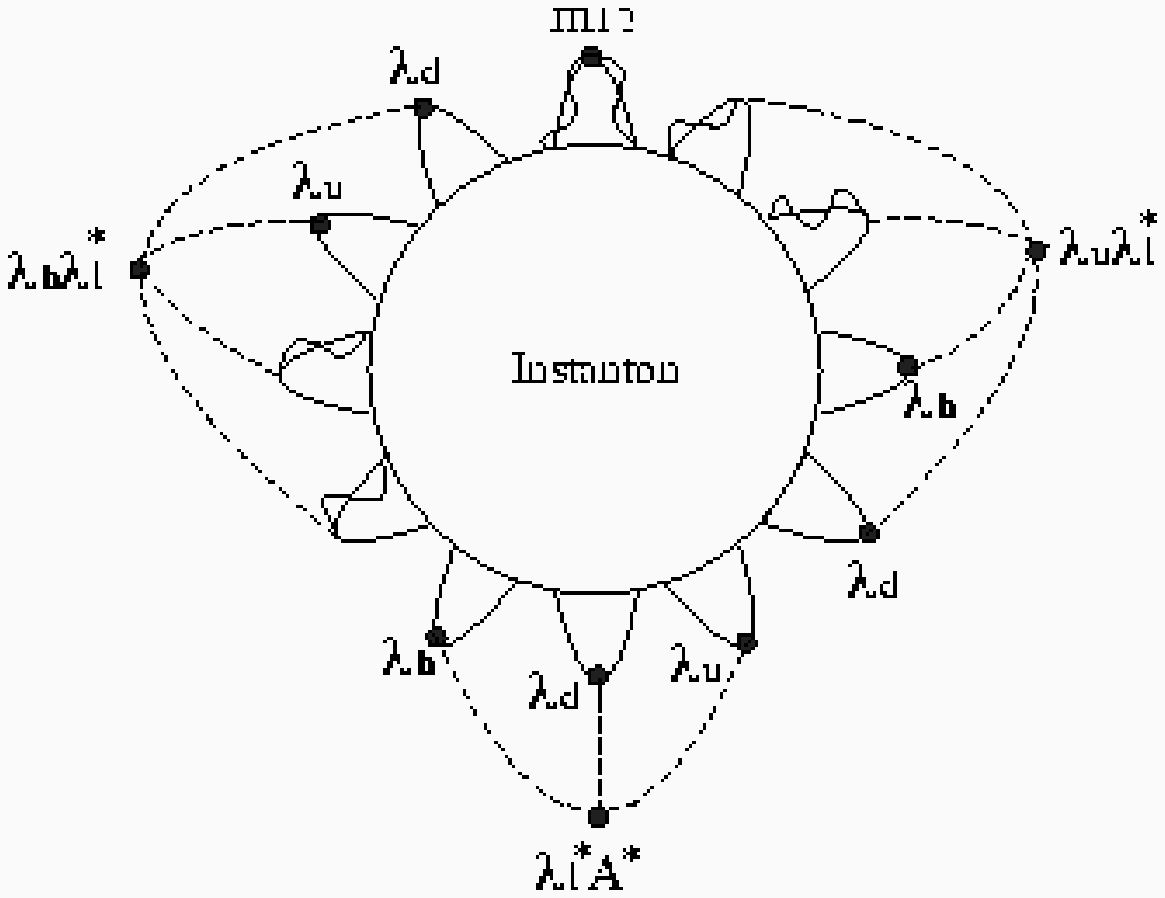, height=9cm}}
\label{fig1}
\newpage
\caption{Instanton graph with the minimal values of $N_S$ and 
the number of loops $L$ for the case of
$N_h = 9$ and $(n,l,k)=(1,1,0)$: $N_S=3/2$ and $L=11$.} 
\vspace{5mm}
\centerline{\epsfig{file=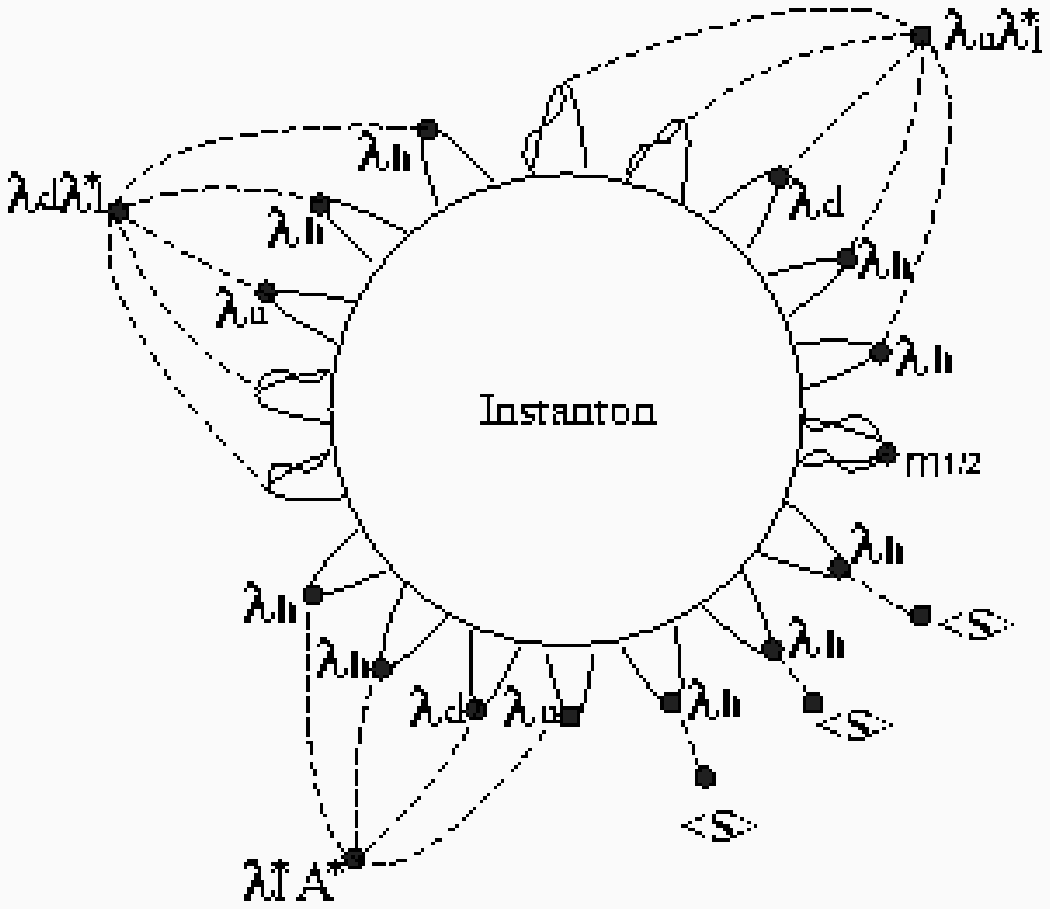, height=9cm}}
\label{fig5}
\vspace{10mm}
\caption{Instanton graph with the same $N_S=3/2$ as Fig. 5 but a larger
number of loops $L=13$ for the case of
$N_h = 9$ and $(n,l,k)=(1,1,0)$.} 
\vspace{5mm}
\centerline{\epsfig{file=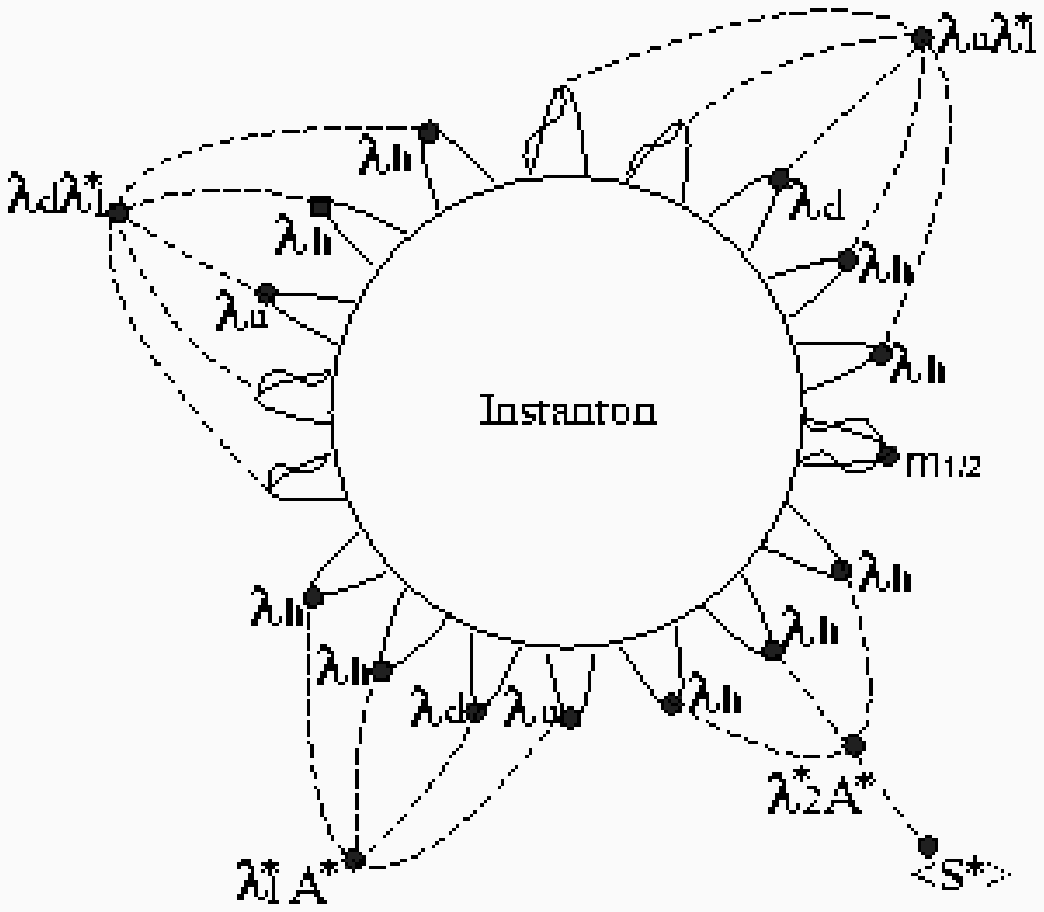, height=9cm}}
\label{fig6}
\end{figure}


\begin{references}
\bibitem{strongCP}
J. E. Kim, \prep{150}{87}{1};
H.-Y. Cheng, \prep{158}{88}{1};
R. D. Peccei, in {\bf CP Violation}, ed. C. Jarlskog (World Scientific,
Singapore, 1989);
R. D. Peccei, hep-ph/9807516.
\bibitem{PQ}
R. D. Peccei and H. R. Quinn, \prl{38}{77}{1440}; \prd{16}{77}{1791};
\bibitem{dine}
M. Dine and N. Seiberg, \npb{273}{86}{109}.
\bibitem{flynn}
J. Flynn and L. Randall, \npb{293}{87}{731}.
\bibitem{georgi}
H. Georgi and L. Randall, \npb{276}{86}{241}.
\bibitem{grav}
H. P. Nilles \prep{110}{84}{1}.
\bibitem{gauge}
M. Dine and A. E. Nelson, \prd{48}{93}{1277};
M. Dine, A. E. Nelson and Y. Shirman, \prd{51}{95}{1362}.
M. Dine, A. E. Nelson, Y. Nir and Y. Shirman, \prd{53}{96}{2658}.
\bibitem{choi}
K. Choi, H. B. Kim, and J. E. Kim, \npb{490}{97}{349}.
\bibitem{munoz}
C. Munoz, private communications.
\end{references}
\end{document}